**Acetylene-Accelerated Alcohol Catalytic CVD Growth of Vertically Aligned Single-Walled Carbon Nanotubes****

*Rong Xiang, Erik Einarsson, Jun Okawa, Yuhei Miyauchi, Shigeo Maruyama**

[*] Prof. S. Maruyama, Mr. R. Xiang, Dr. E. Einarsson, Mr. J. Okawa,
Department of Mechanical Engineering, The University of Tokyo
7-3-1 Hongo, Bunkyo-ku, Tokyo 113-8656, Japan
E-mail: maruyama@photon.t.u-tokyo.ac.jp

Dr. Y. Miyauchi
Institute for Chemical Research, Kyoto University, Uji, Kyoto 611-0011, Japan

[**] Part of this work was financially supported by Grants-in-Aid for Scientific Research (19206024 and 19054003) from the Japan Society for the Promotion of Science, SCOPE (051403009) from the Ministry of Internal Affairs and Communications, NEDO (Japan), and MITI's Innovation Research Project on Nanoelectronics Materials and Structures.

Supporting Information is available on the WWW under http://www.small-journal.com or from the author.

Abstract

Addition of only 1% of acetylene into ethanol was found to enhance the growth rate of single-walled carbon nanotubes (SWNTs) by up to ten times. Since acetylene is a byproduct of the thermal decomposition of ethanol, this suggests an alternative fast reaction pathway to the formation of SWNTs from ethanol via byproducts of decomposition. This accelerated growth, however, only occurred in the presence of ethanol, whereas pure acetylene at the same partial pressure resulted in negligible growth and quickly deactivated the catalyst. The dormant catalyst could be revived by reintroduction of ethanol, indicating that catalyst deactivation is divided into reversible and irreversible stages.

The excellent properties of single-walled carbon nanotubes (SWNTs), particularly chirality-dependent electrical conductivity,[1] make them one of the most exciting materials in nanoscience and nanoengineering. Accordingly, many potential applications of SWNTs have been proposed.[2] After the successful synthesis of random SWNTs by various methods,[3-5] the focus shifted to assembling these one-dimensional tubes into vertically-aligned arrays, which offer at least two intrinsic advantages. First, aligned arrays possess preferable and sometimes unique properties, such as anisotropic electrical and thermal transport, and polarization-dependent optical absorption.[6] These anisotropic materials have the potential for improving performance in existing applications or the development of novel applications, as already demonstrated using multi-walled carbon nanotubes (MWNTs).[7-9] The other intrinsic advantage of an aligned array is that it offers an ideal platform to study the growth mechanism and kinetics, as nearly all the SWNTs have lengths approximately equal to the height of the array.

Historically, the first vertically aligned MWNT array was synthesized in 1996.[10] However, it was not until recently that vertically-aligned SWNT arrays were obtained, first from alcohol catalytic chemical vapor deposition (ACCVD),[11] and followed soon after by many other methods including water-assisted CVD,[12] microwave plasma CVD,[13] and hot-filament CVD.[14] Recent investigations have also improved our understanding of SWNT growth behavior.[15-19] We previously developed an *in situ* optical absorption measurement that allows for convenient real-time measurement of the film thickness.[15] This technique revealed, for the first time, *in situ* growth kinetics of a VA-SWNT array. It provided sub-second resolution and much more direct information to the previously black-box approach to studying the growth process. The *in situ* measurement setup is shown in Fig. 1a, where a laser is passed through the SWNT array, and the thickness is determined from the absorption. Figure 1b shows a typical growth curve obtained from this technique. The SWNTs grow very fast at the beginning but the growth rate decreases with time, effectively stopping within 10 min. The diminishing growth rate can be fitted almost perfectly by an exponential decay, which can be expressed in terms of an initial growth rate and a decay time constant.[20] The detailed influences of CVD parameters, such as temperature and growth pressure, have been reported elsewhere.[21]

Here we investigate the influence of various species on ACCVD synthesis of VA-SWNTs. A small amount of acetylene (approximately 1% partial pressure) was found to accelerate the growth rate by almost ten times, as revealed by a distinct change in the growth rate determined from *in situ* optical absorption. Additionally, the inherent contribution of acetylene, which can be produced by thermal decomposition of ethanol, to ACCVD is also quantitatively investigated.

The influence of various additive species on SWNT growth was determined by introducing different gases into the growth stage of a SWNT array. Using the *in situ* absorption measurement technique was very convenient and straightforward because the effect of additive species is quickly and directly obvious by how they change the growth rate. With this method, the influence of different precursors should be apparent immediately after their introduction, at a known catalyst condition. This gives more reliable results than any *ex situ* methods where catalyst activity may vary among different substrate CVD runs. Several typical examples are presented in Fig. 1c. The ethanol flow rate for all cases was 450 sccm (standard cubic centimeters per minute). When 0.4 sccm of air (< 0.1 % of the ethanol flow) was introduced into the chamber, the growth rate decreased slightly, but was not completely terminated. Similar results were observed for the addition of water ($\leq$ 5 vol%) into the ethanol. This confirms the robustness of our ACCVD; possible leaks and small amounts of impurities in the precursor will not cause complete failure of SWNT formation, but only affect the final SWNT yield.

Addition of 300 sccm of Ar/H$_2$ (3% H$_2$) resulted in neither acceleration nor deceleration of the growth. The synthesis reaction was, however, very sensitive to the addition of even a small amount of acetylene. Figure 1c shows a clear growth enhancement resulting from the addition of only 3.5 sccm acetylene (0.8 vol%). As the derivative of the growth curve gives the growth rate (Fig. 1d), it is clear that, within seconds, the growth rate increased from 0.15 to 1.2 μm/s, an eightfold increase. There was no noticeable change, however, from the addition of a similar amount of ethylene (Fig. 1c) or methane (not shown), two popular carbon sources used for SWNT growth.

Figure 2a shows a typical SEM image of an acetylene-accelerated VA-SWNT array. The SWNTs are well-aligned, and the thickness (~20 μm) agrees well with the estimate from laser absorption. Since the acetylene was introduced when the array had reached 5 μm in height, the upper 5 μm of the aligned SWNT array was produced from ethanol, and the remaining 15 μm primarily from acetylene. The VA-SWNT array shown in Fig. 2b was synthesized in a similar fashion, and a similar growth enhancement was observed. In this case, however, isotope-modified 1,2-$^{13}$C acetylene was used. To characterize the quality of these SWNTs, Raman spectra were obtained from different positions along the cross-section of the SWNT array. For normal acetylene (Fig. 1a), similar spectra were obtained from the upper 1/4, the center, and the lower 1/4 of the array. TEM observation also confirmed that the arrays consist of SWNTs with little amorphous carbon and no MWNTs. As there is no significant difference, in the different regions, it is probable that there was no interruption of growth when acetylene was introduced, and the SWNTs maintained their original chiralities.[22] For the isotope-modified case, however, a shift of all Raman peaks was observed in the lower part of the SWNT array, due to the reduced phonon energy caused by the introduction of the heavier $^{13}$C isotope.[23] The G-band shifted from 1590 ($^{12}$C) to 1544 cm$^{-1}$, which suggests a $^{13}$C content of over 80%. This again illustrates the high reactivity of acetylene.

The first question that arises regarding this acetylene-boosted growth is whether acetylene alone can grow SWNTs at similar low concentrations but in the absence of ethanol. To test this we performed CVD with pure acetylene at the same low pressures as when ethanol was present. The results are presented in Fig. 3a, and are compared with a typical ethanol CVD case. The growth curves suggest some carbon deposition, but the reaction stopped within 20 s. Although Raman spectroscopy confirmed the presence of well-graphitized carbon (spectra not shown), the yield was negligible when compared to conventional ACCVD. This shows that ethanol is necessary to initiate growth, which can later be accelerated by acetylene.

The next natural question which arises is whether or not acetylene can sustain growth initiated by ethanol. This was investigated by introducing the gases in different stages, as shown in Fig. 3b. SWNT growth was initiated by ethanol (blue line), and maintained for 30 seconds. The ethanol was then stopped and the chamber evacuated, followed immediately by introduction of acetylene (red line). The acetylene initiated a very small growth spurt, but the catalyst quickly became inactive. This indicates that ethanol is necessary throughout the entire growth process, both for cap formation and to sustain growth. Interestingly, reintroduction of ethanol in addition to the already-present acetylene (purple line in Fig. 3b) *recovered* the activity of the acetylene-poisoned catalysts, despite showing no signs of activity for more than 60 seconds. Longer exposure to pure acetylene was found to reduce the degree to which the catalyst could be recovered by reintroduction of ethanol, with recovery being impossible after more than five minutes.

The high reactivity of acetylene is not surprising, but it is interesting that in this case it cannot work alone, and that ethanol can recover the deactivated catalyst. One possible mechanism causing this catalyst deactivation by acetylene and re-activation by ethanol is shown in the inset of Fig. 3b. It is known that hydrocarbons produce much more soot than alcohols,[24, 25] which can likely coat the catalyst and prevent further feedstock supply. At this point the catalyst is likely only inactive rather than completely deactivated, and could be recovered by removing this amorphous carbon coating. However, after prolonged exposure to such a highly-reactive carbon source and a high-temperature environment, this amorphous layer may form a more stable graphitic layer, or form a carbide particle, after which the catalyst would become irreversibly deactivated, and could not be recovered by exposure to ethanol. According to our experience, the time needed for this irreversible deactivation is several minutes. We suspect this process because at our growth conditions, ethanol is believed to efficiently remove amorphous carbon,[24] but does not damage graphitic carbon, e.g., SWNTs.[20] Additionally, as long as there is no carbon source available to the catalyst, e.g., in an $Ar/H_2$ atmosphere, the catalyst can be kept active at CVD temperatures for 60 minutes or longer. This effectively rules out catalyst-substrate interactions or aggregation of metal particles[26] as potential deactivation mechanisms.

The concentration of acetylene in the present work is approximately 10 Pa, which is much lower than most previous studies. However, when changing the flow rate of acetylene, we found only 0.35 sccm acetylene (1 Pa) was needed to double the growth rate of 1.3 kPa of ethanol. Quantitatively, the rate of collisions between ethanol molecules and catalyst nanoparticles ($k_1$) is at least 1000 higher than those involving acetylene molecules ($k_2$), hence acetylene is at least 1000 times more active than ethanol. The growth enhancement factor, defined as ($R_{ethanol+acetylene}/R_{ethanol}$)-1, is shown in Fig. 4a, and increases almost linearly at very low acetylene partial-pressure. This indicates that SWNT growth in conventional ACCVD is limited by surface reactions rather than carbon diffusion in/on the metal nanoparticle, as the catalyst clearly have potential to produces SWNTs at a much faster rate.

As most feedstock species undergo some degree of thermal decomposition before reaching the catalyst, understanding the reaction pathway and the actual reacting species are important in improving understanding of the CVD processes and control over the final SWNT product.

The primary byproducts of ethanol thermal decomposition at 800 °C are ethylene and water (see supporting information)[27, 28], but small amounts of other chemical species are also produced, including acetylene. Since acetylene has shown to be highly active, even in very low concentration, the apparent activity of ethanol may come partly from acetylene, thus we must consider is a third possible reaction pathway, with rate $k_3$, in Fig. 4a. Also, although ethylene is not as active as acetylene the concentration of ethylene is around two orders of magnitude higher than acetylene, thus SWNTs grown from ethylene may not be negligible. Further work is needed to determine quantitatively the contribution by these indirect pathways to SWNT synthesis, although we can determine qualitatively whether or not ethanol must decompose into acetylene or ethylene in order to grow SWNTs.

Since ethanol has an asymmetric molecular structure, labeling the two carbon atoms by using isotopes makes it possible to qualitatively determine the contribution of acetylene or ethylene to SWNT synthesis. We synthesized SWNTs using normal 1,2-$^{12}$C ethanol and three isotope-modified versions: 1-$^{13}$C, 2-$^{13}$C, and 1,2-$^{13}$C ethanol. Using a no-flow condition,[18] where a fixed amount of ethanol is sealed in the chamber and therefore has sufficient time to decompose before forming SWNTs, Raman spectra indicate that 2-C (the carbon atom opposite the OH group) contributes slightly more than the 1-C to the final product (Fig. 4b). Both acetylene and ethylene are produced by breaking the C-O bond in ethanol, and are perfectly symmetric in structure thus are expected to have equal contribution from the 1-$^{13}$C and 2-$^{13}$C. This result indicates that ethanol is directly reacting with the catalyst in ACCVD, and is not simply the starting material. The contribution from ethanol increases when using the standard constant-flow case, and the unequal contribution of two different carbon atoms becomes even larger (~70% 2-C incorporation into the final product)[29] using a fast-flow condition (*e.g.*, by injecting through a nozzle). This is expected, as the faster flow reduces the degree of thermal decomposition prior to arrival at the catalyst, and confirms that the ratio of the various chemical pathways to SWNT formation is altered by the extent of ethanol decomposition. We also find that in the no-flow case, when ethanol is thoroughly decomposed, the SWNTs contain more amorphous carbon than those formed more directly from ethanol. This experiment not only provides us quantitative growth kinetics for ethanol and acetylene feedstock, but also emphasizes the critical role of ethanol. Improved understanding may offer a better way to control the quality and purity of as-grown SWNTs. More detailed work is in progress.

In summary, we investigated acetylene-acceleration of ACCVD, in which a small concentration of acetylene in addition to ethanol was found to significantly enhance the growth rate of SWNTs. The very high activity of acetylene (estimated to be 1000 times that of ethanol) also suggests a possible fast chemical pathway from ethanol to SWNT formation. However, this fast growth only occurred in the presence of ethanol; pure acetylene at the same partial pressure resulted in negligible growth, deactivating the catalyst in a few seconds. However, these dormant catalyst particles could be reactivated by reintroduction of ethanol. This indicates that catalyst deactivation is divided into reversible and irreversible stages, and also shows the ability of ethanol to preserve catalyst activity throughout the synthesis process. As thermal decomposition is common to almost all CVD systems, it is important to understand quantitatively which species are actually present and their respective roles in contributing to SWNT growth. Further investigation in this direction should not only result in a more complete understanding of the synthesis and catalyst deactivation mechanisms, but also better control of the entire synthesis process.

**Experimental Section**

Vertically aligned SWNTs were synthesized at 800 °C using ethanol as a carbon source and a Co/Mo combination as a catalyst.[30] The catalyst was dip-coated onto quartz using a two-step procedure, as described in our previous work.[21] The substrate was then annealed in air at 400 °C for 3 min before heated to 800 °C under a 300 sccm $Ar/H_2$ flow (3% $H_2$, Ar balance) at a pressure of 40 kPa. Upon reaching the growth temperature, the $Ar/H_2$ flow was stopped and 450 sccm of ethanol was introduced at 1.3 kPa to start SWNT growth. Additive species, typically acetylene at 0.3 to 14 sccm, were introduced 30 s after the introduction of ethanol. The growth behavior was monitored *in situ* by passing a 488 nm laser through the quartz substrate, as shown in Fig. 1a, and the thickness of the SWNT array was determined from the real-time absorption measurement.[15] The as-grown samples were characterized by SEM (JEOL 7000F, operated at 3 kV), TEM (JEOL 2000EX, operated at 120 kV), and Raman spectroscopy (488 nm excitation).

[1]     S. Iijima, T. Ichihashi, *Nature* **1993**, *363*, 603.
[2]     Carbon Nanotubes: Advanced Topics in the Synthesis, Properties and Applications (Eds.: A. Jorio, M. S. Dresselhaus, G. Dresselhaus), *Springer*, *2007*.
[3]     A. Thess, R. Lee, P. Nikolaev, H. J. Dai, P. Petit, J. Robert, C. H. Xu, Y. H. Lee, S. G. Kim, A. G. Rinzler, D. T. Colbert, G. E. Scuseria, D. Tomanek, J. E. Fischer, R. E. Smalley, *Science* **1996**, *273*, 483.
[4]     C. Journet, W. K. Maser, P. Bernier, A. Loiseau, M. L. de la Chapelle, S. Lefrant, P. Deniard, R. Lee, J. E. Fischer, *Nature* **1997**, *388*, 756.
[5]     H. J. Dai, A. G. Rinzler, P. Nikolaev, A. Thess, D. T. Colbert, R. E. Smalley, *Chem. Phys. Lett.* **1996**, *260*, 471.
[6]     Y. Murakami, E. Einarsson, T. Edamura, S. Maruyama, *Phys. Rev. Lett.* **2005**, *94*, 087402.
[7]     H. Huang, C. H. Liu, Y. Wu, S. S. Fan, *Adv. Mater.* **2005**, *17*, 1652.
[8]     K. L. Jiang, Q. Q. Li, S. S. Fan, *Nature* **2002**, *419*, 801.
[9]     A. Y. Cao, P. L. Dickrell, W. G. Sawyer, M. N. Ghasemi-Nejhad, P. M. Ajayan, *Science* **2005**, *310*, 1307.
[10]    W. Z. Li, S. S. Xie, L. X. Qian, B. H. Chang, B. S. Zou, W. Y. Zhou, R. A. Zhao, G. Wang, *Science* **1996**, *274*, 1701.
[11]    Y. Murakami, S. Chiashi, Y. Miyauchi, M. H. Hu, M. Ogura, T. Okubo, S. Maruyama, *Chem. Phys. Lett.* **2004**, *385*, 298.
[12]    K. Hata, D. N. Futaba, K. Mizuno, T. Namai, M. Yumura, S. Iijima, *Science* **2004**, *306*, 1362.
[13]    G. F. Zhong, T. Iwasaki, K. Honda, Y. Furukawa, I. Ohdomari, H. Kawarada, *Chem. Vapor Depos.* **2005**, *11*, 127.
[14]    G. Y. Zhang, D. Mann, L. Zhang, A. Javey, Y. M. Li, E. Yenilmez, Q. Wang, J. P. McVittie, Y. Nishi, J. Gibbons, H. J. Dai, *Proc. Natl. Acad. Sci. U.S.A.* **2005**, *102*, 16141.
[15]    S. Maruyama, E. Einarsson, Y. Murakami, T. Edamura, *Chem. Phys. Lett.* **2005**, *403*, 320.
[16]    D. N. Futaba, K. Hata, T. Yamada, K. Mizuno, M. Yumura, S. Iijima, *Phys. Rev. Lett.* **2005**, *95*.
[17]    T. Iwasaki, G. F. Zhong, T. Aikawa, T. Yoshida, H. Kawarada, *J. Phys. Chem. B* **2005**, *109*, 19556.
[18]    R. Xiang, Z. Zhang, K. Ogura, J. Okawa, E. Einarsson, Y. Miyauchi, J. Shiomi, S. Maruyama, *Jpn. J. Appl. Phys.* **2008**, *47*, 1971.
[19]    R. Xiang, Z. Yang, Q. Zhang, G. H. Luo, W. Z. Qian, F. Wei, M. Kadowaki, S. Maruyama, *J. Phys. Chem. C* **2008**, *112*, 4892.
[20]    E. Einarsson, Y. Murakami, M. Kadowaki, S. Maruyama, *Carbon* **2008**, *46*, 923.
[21]    E. Einarsson, M. Kadowaki, K. Ogura, J. Okawa, R. Xiang, Z. Zhang, Y. Yamamoto, Y. Ikuhara, S. Maruyama, *J. Nanosci. Nanotechnol.* **2008**, *8*, 6093.
[22]    Y. H. Wang, M. J. Kim, H. W. Shan, C. Kittrell, H. Fan, L. M. Ericson, W. F. Hwang, S. Arepalli, R. H. Hauge, R. E. Smalley, *Nano Lett.* **2005**, *5*, 997.
[23]    Y. Miyauchi, S. Maruyama, *Phys. Rev. B* **2006**, *74*, 35415.
[24]    S. Maruyama, R. Kojima, Y. Miyauchi, S. Chiashi, M. Kohno, *Chem. Phys. Lett.* **2002**, *360*, 229.
[25]    J. Warnatz, U. Maas, R. W. Dibble, Combustion: Physical and Chemical Fundamentals, Modeling and Simulation, Experiments, Pollutant Formation, third edn., Springer, Berlin, **2001**, 257.

[26] Jung Y J, Wei B Q, Vajtai R, Ajayan P M, Homma Y, Prabhakaran K, Ogino T, *Nano Lett.* **2003**, *3*, 561.
[27] G. Rotzoll, *Journal of Analytical and Applied Pyrolysis* **1985**, *9*, 43.
[28] H. Sugime, S. Noda, S. Maruyama, Y. Yamaguchi, *Carbon* **2008**, DOI: 10.1016/j.carbon.2008.10.001.
[29] S. Maruyama, Y. Miyauchi, *Electronic Properties of Novel Nanostructure (AIP Conf. Proc. 786)* **2005**, 100.
[30] M. H. Hu, Y. Murakami, M. Ogura, S. Maruyama, T. Okubo, *J. Catal.* **2004**, *225*, 230.

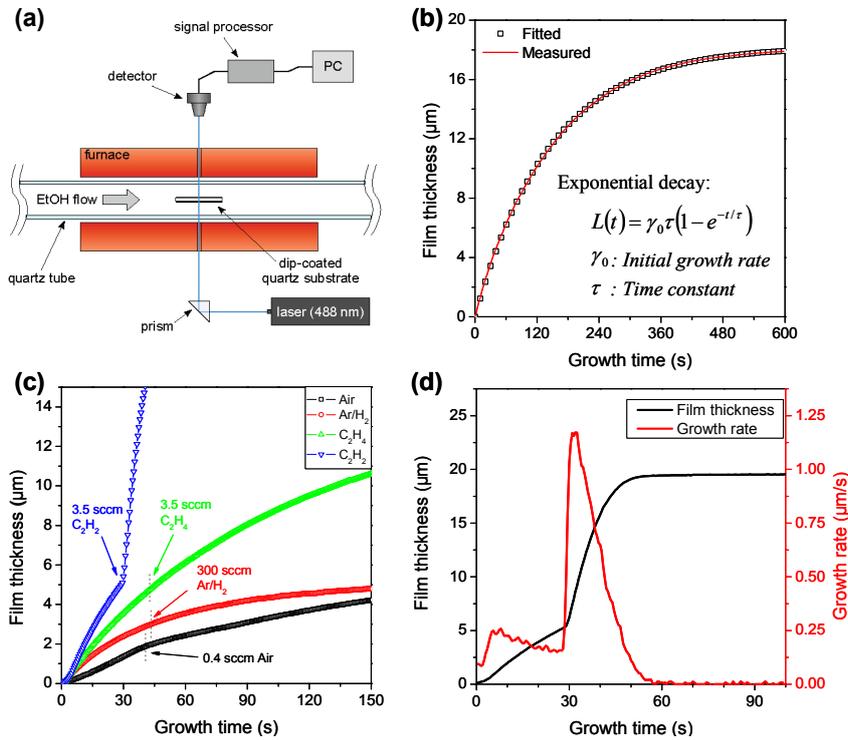

**Figure 1.** (a) Schematic of *in situ* optical absorption measurement; (b) a typical growth curve obtained from optical absorption, which can be fitted by an exponential decay; (c) influence of various additive species on the growth of aligned SWNTs; (d) VA-SWNT film thickness (black line) and growth rate (red line) showing acceleration of SWNT growth by addition of acetylene.

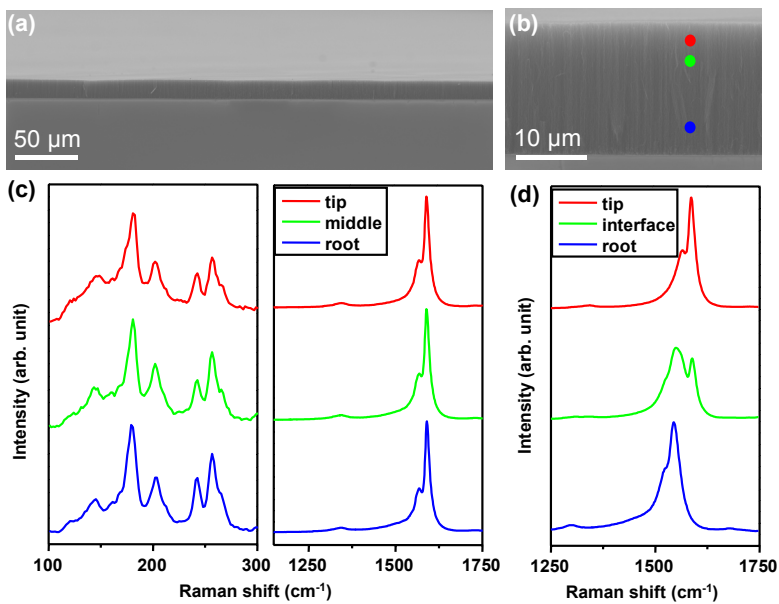

**Figure 2.** SEM images of as-grown SWNT arrays from ACCVD boosted by (a) $^{12}$C acetylene and (b) $^{13}$C acetylene; (c) Raman spectra taken from different positions of an acetylene-accelerated SWNT array, showing almost identical G-bands and radial breathing mode peaks; (d) G-band of a SWNT array accelerated by $^{13}$C acetylene.

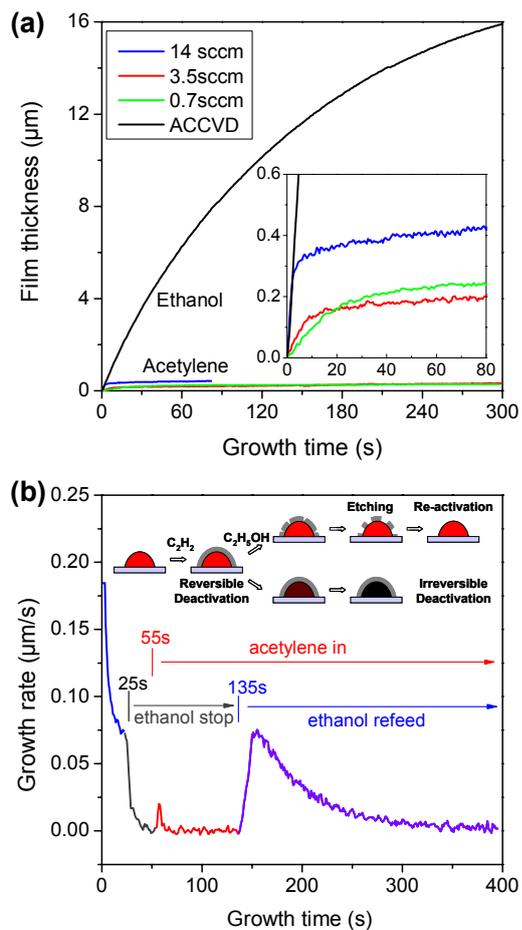

**Figure 3.** (a) Growth curves of CVD with pure acetylene at different flow rates showing almost negligible yield and fast poisoning of the catalyst; (b) CVD started with ethanol but continued with pure acetylene, showing similar fast catalyst deactivation, but the activities of those apparently deactivated catalysts could be recovered by ethanol. A possible schematic presentation of this deactivation and re-activation process is shown as an inset.

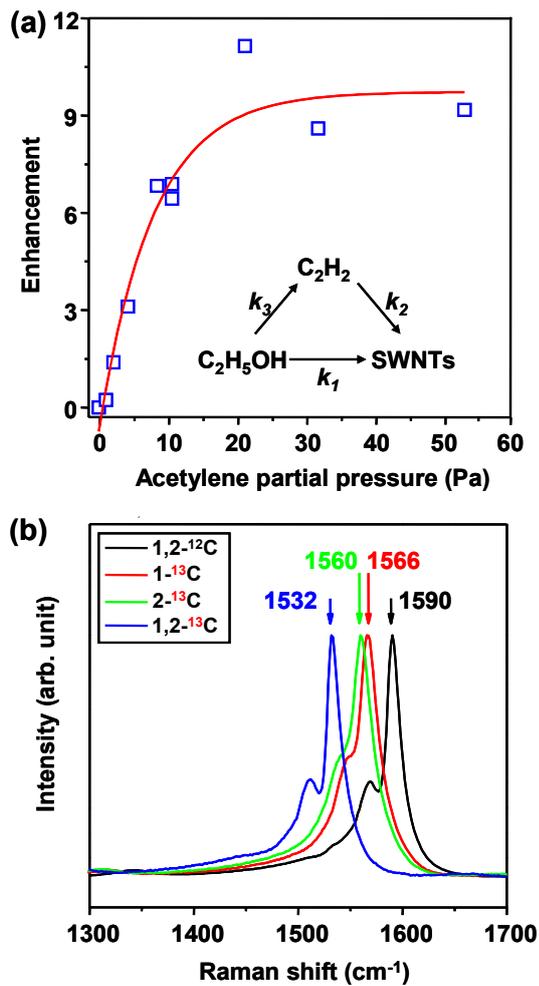

**Figure 4.** (a) Influence of the acetylene flow rate on the degree of growth enhancement, as only about 0.1% of acetylene can reach similar growth of 450 sccm of ethanol; (b) Raman spectra of SWNTs grown from four types of ethanol with different isotope configurations, showing that the second carbon contributes more to SWNT formation. This confirms that not all SWNTs form from decomposed ethanol.

We find single-walled carbon nanotubes (SWNTs) synthesis from ethanol is significantly accelerated by the addition of a small amount of acetylene. In the absence of ethanol, however, catalyst activity quickly diminishes, and negligible growth occurs, but reintroduction of ethanol can recover the catalyst activity. These results show the importance of secondary precursors in SWNT formation.

TOC Keyword: aligned single walled carbon nanotube array, *in situ* absorption, acetylene, CVD

R. Xiang, E. Einarsson, J. Okawa, Y. Miyauchi, S. Maruyama

**Acetylene-Accelerated Alcohol Catalytic CVD Growth of Vertically Aligned Single-Walled Carbon Nanotubes**

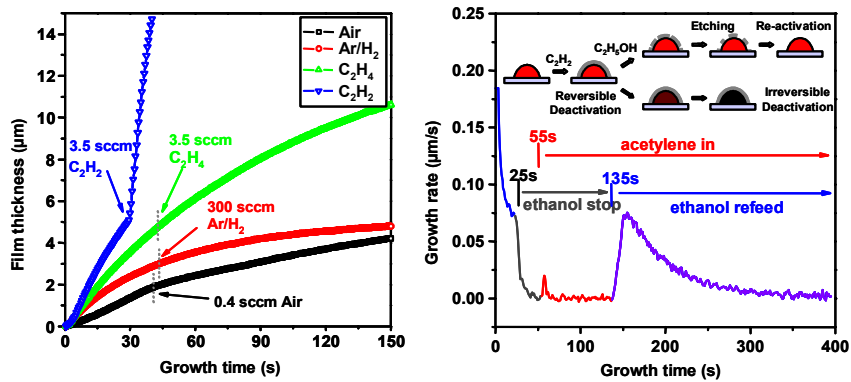

**Supporting Information**

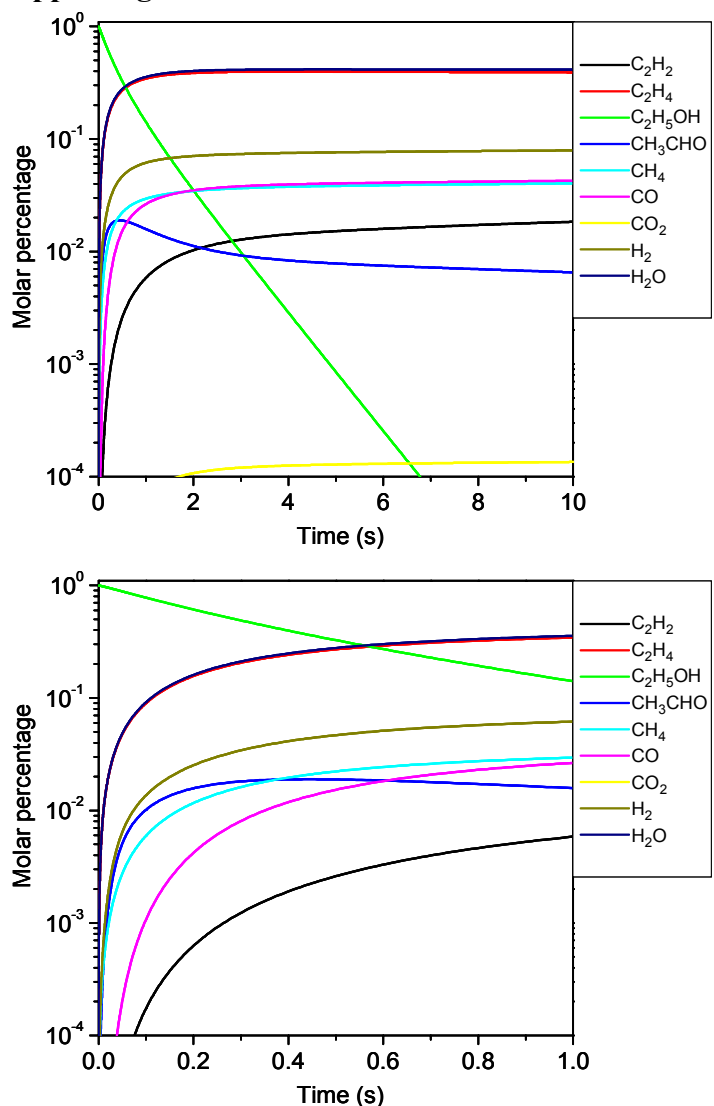

**Figure S1.** Calculated time-dependent gas composition due to thermal decompostion of ethanol at constant temperature of 800 °C and constant pressure of 1.3 kPa. This indicates that in typical ACCVD, ethanol decomposes within one second, and that acetylene is one of the byproducts. Calculation was performed using the software package SENKIN, based on the kinetic model proposed by N. M. Marinov.[1]

[1] N. M. Marinov, *Int. J. Chem. Kinet.* **1999**, 31,183-220.